\documentclass[journal]{IEEEtran}

\ifCLASSINFOpdf
\else
   \usepackage[dvips]{graphicx}
\fi
\usepackage{url}

\usepackage{graphicx}
\usepackage{amsmath,amssymb,graphicx}
\usepackage[]{algorithm2e}
\usepackage{tabularx}
\usepackage{wasysym}
\usepackage{color,soul}
\usepackage{cite}
\usepackage{balance}
\usepackage{booktabs}
\usepackage{multirow}
\usepackage{siunitx}

\usepackage{makecell}
\usepackage{caption}
\usepackage[font=scriptsize]{subfig}
\newcommand{\subparagraph}{}
\usepackage{titlesec}
\usepackage{textgreek}

\usepackage{soul}
\soulregister\cite7 %
\soulregister\citep7 %
\soulregister\citet7 %
\soulregister\ref7 %
\soulregister\pageref7 %
\newcommand{\ZQHL}[1]{{#1}} %

\begin{document}

\setlength{\abovedisplayskip}{2.5pt}
\setlength{\belowdisplayskip}{2.5pt}

\titlespacing*{\section}{0pt}{2.ex plus 0ex minus 0ex}{0.5ex plus 0ex minus 0ex}
\titlespacing*{\subsection}{0pt}{2.ex plus 0ex minus 0ex}{0.5ex plus 0ex minus 0ex}

\title{On The Compensation Between Magnitude and Phase in Speech Separation}

\author{Zhong-Qiu Wang, Gordon Wichern, and Jonathan Le Roux
\thanks{Manuscript received on Jun. 17, 2021; revised on Sep. 14, 2021.}
\thanks{Z.-Q. Wang, G. Wichern, and J. Le Roux are with Mitsubishi Electric Research Laboratories, Cambridge, MA 02139, USA (e-mail: wang.zhongqiu41@gmail.com, \{wichern,leroux\}@merl.com).}}

\markboth{IEEE Signal Processing Letters}
{Shell \MakeLowercase{\textit{et al.}}: Bare Demo of IEEEtran.cls for IEEE Journals}

\maketitle

\begin{abstract}

Deep neural network (DNN) based end-to-end optimization in the complex time-frequency (T-F) domain or time domain has shown considerable potential in monaural speech separation.
Many recent studies optimize loss functions defined solely in the time or complex domain, without including a loss on magnitude.
Although such loss functions typically produce better scores if the evaluation metrics are objective time-domain metrics, they however produce worse scores on speech quality and intelligibility metrics and usually lead to worse speech recognition performance, compared with including a loss on magnitude.
While this phenomenon has been experimentally observed by many studies, it is often not accurately explained and there lacks a thorough understanding on its fundamental cause.
This paper provides a novel view from the perspective of the implicit compensation between estimated magnitude and phase.
Analytical results based on monaural speech separation and robust automatic speech recognition (ASR) tasks in noisy-reverberant conditions support %
the validity of our view.

\end{abstract}

\vspace{-0.125cm}
\begin{IEEEkeywords}
End-to-end optimization, speech enhancement, speaker separation, phase estimatiion, deep learning.

\end{IEEEkeywords}

\IEEEpeerreviewmaketitle

\section{Introduction}

\IEEEPARstart{D}{eep} learning has elevated the performance of speech separation in the past decade.
Early DNN based approaches operated in the magnitude domain, where DNNs were trained based on magnitude features to predict target magnitudes directly or via estimating T-F masks, and the mixture phase was used for signal re-synthesis \cite{WDLreview}.
Popular T-F masks 
include the ideal binary/ratio mask \cite{WDLreview}, ideal amplitude mask (IAM) \cite{WYXtrainingtargets}, and phase-sensitive mask (PSM) \cite{Erdogan2015}.
To obtain better phase for re-synthesis, subsequent studies estimated it by using DNN-estimated magnitudes to drive iterative phase reconstruction (IPR) algorithms \cite{Han2015, WZQe2eMISI2018, WZQtrigonometric2019, Zhao2019}.
Building upon deep learning based end-to-end optimization, recent studies implicitly estimate phase by predicting the real and imaginary (RI) components of target speech from the mixture RI components \cite{Williamson2016, Fu2017, tan2020, Wang2020a, Wang2020b, Wang2020d, Wang2020c}, or predicting target waveforms from the mixture waveform \cite{Pascual2017, Rethage2018, Luo2019, Pandey2019, Defossez2020}.
This end-to-end approach has shown large improvements over magnitude-domain approaches.
Many studies along this line define their loss functions only in the time domain to optimize for example SI-SDR \cite{Luo2019}, or in the complex domain to minimize an $L_p$-norm distance between predicted and target RI components \cite{tan2020, Liu2019}.
Some studies add a loss on the magnitude of the predicted RI components or waveforms \cite{Fu2017, WZQe2eMISI2018, Isik2020, Wisdom2018, Pandey2019, Defossez2020, Wang2020a, Wang2020b, Wang2020d, Wang2020c, Li2021, Braun2021, Aroudi2021, Sawata2021}.
This loss is reported to produce clear improvements in speech quality, intelligibility, and ASR scores with slightly worse SI-SDR results \cite{Wang2020a, Wang2020b, Wang2020d, Wang2020c}.
Although this phenomenon has been experimentally observed in many studies, its fundamental cause is often misinterpreted or not thoroughly analyzed.

Our study provides a novel explanation to this observation.
Our insight is that, since phase is always difficult to estimate accurately, if the loss is defined solely in the complex or time domain, the magnitude of the estimated speech will tend to compensate for an inaccurate phase estimate, leading to a less accurate magnitude compared with the one obtained by including a magnitude loss, or, alternatively, training a magnitude-domain model for direct magnitude estimation.
We shall point out that many techniques in this paper have been proposed before.
Our contribution is a novel view on why they work well and under what conditions they would work less well.
Such a view can facilitate understanding and guide algorithmic design in speech separation.

\section{End-to-End Speech Separation}\label{e2edescription}

\noindent We review two popular end-to-end approaches for speech separation.
Given a monaural mixture, the physical model relating the mixture $y$, target $s$, and non-target signals $v$ can be formulated in the time domain as
$y[n] = s[n]+v[n]$,
where $n$ indexes discrete time.
In the short-time Fourier transform (STFT) domain, the physical model is formulated as
\begin{align} 
	Y(t,f) &= S(t,f)+V(t,f), \label{eq:phymodel_freq}
\end{align}
where $Y$, $S$, and $V$ respectively denote the STFT spectra of $y$, $s$, and $v$, and $t$ and $f$ index time and frequency.

\subsection{Complex-domain Separation}\label{complex}

Complex spectral mapping \cite{Williamson2016, Fu2017, tan2020, Wang2020b, Wang2020a} predicts target RI components from the mixture RI components, simultaneously modeling magnitude and phase.
A typical loss is
\begin{align}\label{riloss}
\mathcal{L}_{\text{RI}} &= \| \hat{R} - \text{Real}(S)\|_1 + \| \hat{I} - \text{Imag}(S)\|_1,
\end{align}
where $\hat{R}$ and $\hat{I}$ are the predicted RI components, $\text{Real}(\cdot)$ and $\text{Imag}(\cdot)$ extract RI components, and $\|\cdotp\|_1$ computes the $L_1$ norm.
The separation result is $\hat{S}=\hat{R}+j\hat{I}$, where $j$ is the imaginary unit.
Inverse STFT (iSTFT) is then applied for signal re-synthesis.
In \cite{Wang2020a, Wang2020b, Wang2020d, Wang2020c}, it is suggested that $\text{STFT}(\text{iSTFT}(\hat{S}))$ is very close to $\hat{S}$, meaning that the magnitude and phase of the estimated complex spectrogram produced by complex spectral mapping are almost consistent with each other.

A magnitude loss can be added \cite{Wang2020a, Wang2020b, Wang2020d, Wang2020c}:
\begin{align}\label{ri+magloss}
\mathcal{L}_{\text{RI+Mag}} &= 
\mathcal{L}_{\text{RI}} + \| |\hat{R}+j\hat{I}| - |S|\|_1,
\end{align}
where $|\cdot|$ extracts magnitude.
One can also train through iSTFT and define the loss in the time domain \cite{WZQe2eMISI2018, WZQtrigonometric2019, Liu2019}:
\begin{align}\label{ri-istftloss}
\mathcal{L}_{\text{RI-iSTFT}} &= 
\| \text{iSTFT}(\hat{S}) - s \|_1.
\end{align}
Note that $\text{iSTFT}(\hat{S})$ is the final signal listened to by end users and used for metric computation.
To improve its magnitude, a magnitude loss can be included \cite{Isik2020, Wisdom2018,  Li2021, Braun2021}:
\begin{align}\label{ri+istft+magloss}
\footnotesize{\mathcal{L}_{\text{RI-iSTFT+Mag}}} &= 
\footnotesize{\mathcal{L}_{\text{RI-iSTFT}}} + \big\| |\text{STFT}\big(\text{iSTFT}\big(\hat{S}\big)\big)| - |S| \big\|_1.
\end{align}
\ZQHL{An alternative computes the magnitude loss before iSTFT:
\begin{align}\label{magloss+ri+istft}
\footnotesize{\mathcal{L}_{\text{Mag+RI-iSTFT}}} &= 
\| |\hat{R}+j\hat{I}| - |S|\|_1 + \footnotesize{\mathcal{L}_{\text{RI-iSTFT}}}.
\end{align}} %
Some studies apply a power compression on the predicted magnitude before computing the loss \cite{Wisdom2018, Li2021power}, use a weight between the time- and magnitude-domain loss, or define the loss on magnitude features \cite{G.Germain2019, Manocha2021} or on multi-resolution magnitudes \cite{Defossez2020}.
They are out of the scope of this paper.

\subsection{Time-domain Separation}\label{time}

Time-domain approaches predict the target waveform directly from the mixture waveform \cite{Pascual2017, Stoller2018, Rethage2018, Luo2019, Pandey2019, Defossez2020}. They implicitly model magnitude and phase through end-to-end optimization. The loss is typically defined solely in the time domain, in the form of mean absolute/square error (or their log-compressed versions and SI-SDR \cite{LeRoux2018a}) as
\begin{align}\label{wavloss}
\mathcal{L}_{\text{Wav}} &= \| \hat{s} - s \|_1,
\end{align}
where $\hat{s}$ denotes the predicted waveform.
Later studies \cite{Pandey2019, Defossez2020} incorporate a magnitude-domain loss
\begin{align}\label{wav+magloss}
\mathcal{L}_{\text{Wav+Mag}} &= \mathcal{L}_{\text{Wav}} + \| |\text{STFT}(\hat{s})| - |S| \|_1.
\end{align}

\section{Compensation Between Magnitude and Phase}\label{compensationdescription}

\noindent This section describes the compensation problem, and loss functions that lead to better magnitude or phase estimation. 

\subsection{The Compensation Problem}\label{compensationproblem}

\begin{figure}
  \centering  
  \includegraphics[width=6cm]{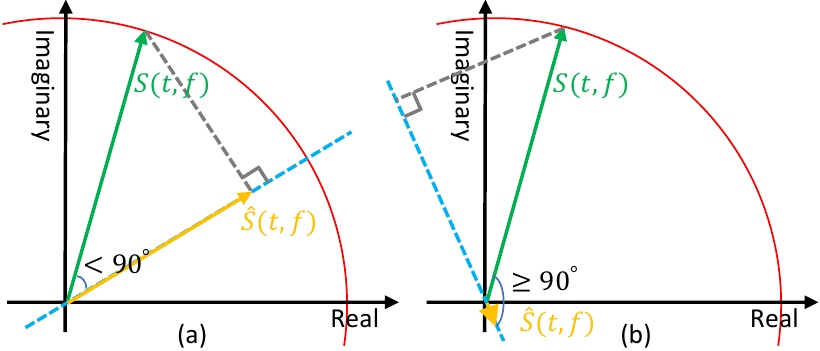}
  \vspace{-0.1cm} \\
  \caption{\scriptsize Complex-plane illustration of magnitude-phase compensation.}
  \vspace{-0.6cm}
  \label{compensationfigure}
\end{figure}

Training a model using (\ref{riloss}) or (\ref{wavloss}) essentially tries to find an estimated speech $\hat{S}(t,f)$ that can approximate clean speech $S(t,f)$ at each T-F unit.
See Fig.~\ref{compensationfigure}(a) for an illustration.
Because phase is generally difficult to estimate, $\angle \hat{S}(t,f)$ is typically very different from $\angle S(t,f)$, especially in T-F units with low signal-to-noise ratio (SNR).
In such cases, the closest approximation of $S(t,f)$ is the projection of $S(t,f)$ onto the direction determined by $\angle \hat{S}(t,f)$.
This approximation, however, is incapable of recovering the clean magnitude, and the difference between the two becomes larger as $\angle \hat{S}(t,f)$ gets away from $\angle S(t,f)$.
If $\angle \hat{S}(t,f)$ is more than $\pi/2$ away from $\angle S(t,f)$, the best approximation would lead to a zero magnitude, as illustrated in Fig.~\ref{compensationfigure}(b).

By including a loss on magnitude in (\ref{ri+magloss}) or (\ref{wav+magloss}), the DNN prediction is encouraged to find a balance between complex- and magnitude-domain approximations.
This balance explains why adding a loss on magnitude leads to better perceptual evaluation of speech quality (PESQ) \cite{Rix2001} and extended short-time objective intelligibility (eSTOI) \cite{Taal2011} scores, because they favor an estimated target time-domain signal that has an accurate magnitude, rather than a compensated, less accurate one.
We point out that eSTOI and STOI \cite{Taal2011} only look at the magnitude envelope of the predicted signal, and
PESQ \cite{Rix2001} first time-aligns the predicted signal with the reference segment-wisely, forgiving any segmental time delays, and then looks at their short-time Bark-scale power spectra.
This balance also suggests that the degradation in time-domain metrics, such as SI-SDR \cite{LeRoux2018a}, is due to the magnitude loss pushing $\hat{S}(t,f)$ away from the closest approximation of $S(t,f)$ along the direction of $\angle \hat{S}(t,f)$.
Indeed, SI-SDR \cite{LeRoux2018a} measures the quality of time-domain sample-level predictions, and hence favors estimated speech with a magnitude spectrogram that compensates for its inaccurate phase spectrogram.

\ZQHL{This compensation view is motivated by the PSM (defined as $|S|/|Y|\text{cos}(\angle S-\angle Y)$) \cite{Erdogan2015}, proposed to find the magnitude closest to $S(t,f)$ along the direction of $\angle Y(t,f)$ for magnitude-domain speech enhancement.
The PSM explicitly considers the compensation so that the SNR of the target estimate can be maximized when $\angle Y$ is used for signal re-synthesis.
Our key contribution is to show that this compensation problem, often neglected, implicitly exists in many end-to-end approaches, which usually improve upon the mixture phase but still cannot perfectly reconstruct the clean phase.
Our study is also the first to %
comprehensively analyze the consequences of this compensation problem on multiple popular evaluation metrics, and emphasizes 
the effectiveness of including a magnitude loss. %
Such an analysis does not exist in \cite{Erdogan2015} and other previous works. %
}

\subsection{Magnitude Spectrogram Approximation}\label{magnitude}

This compensation view suggests that, in cases where we only need a good estimated magnitude and do not have to estimate or leverage phase, it may be better not modelling magnitude and phase simultaneously.
One such scenario is robust ASR based on monaural speech enhancement \cite{WPD2019}, where the recognition model typically only considers magnitude features.
In such cases, direct target magnitude spectrogram approximation (MSA) \cite{Erdogan2015} would likely lead to better magnitude estimation and produce better performance, 
as the resulting magnitude is not a compensated one.
The typical loss function is
\begin{align}\label{MSA}
\mathcal{L}_{\text{MSA}} &= \| \hat{M} - |S|\|_1,
\end{align}
where $\hat{M}$ denotes estimated magnitude. Eq. (\ref{MSA}) can be considered as a teacher-forcing technique \cite{Courville2016, Zhao2019}, assuming that the estimated speech has the same phase as target speech, i.e.,
\begin{align}\label{MSAoraclephase}
\mathcal{L}_{\text{MSA}} &= \| \hat{M}e^{j\angle S} - |S|e^{j\angle S}\|_1.
\end{align}
This way, the implicit compensation between magnitude and phase is avoided, because the best approximation of $S(t,f)$ along the direction of $\angle S(t,f)$ is $|S(t,f)|$.
One potential issue with MSA is that when signal re-synthesis is needed, the mixture phase is typically used together with the estimated magnitude.
This usually leads to sub-optimal SI-SDR, as the best approximation of $S(t,f)$ along $\angle Y(t,f)$ should have a compensated magnitude.
In addition, due to the phase inconsistency issue \cite{Griffin1984a, LeRoux2008, Wisdom2018, WZQtrigonometric2019}, the magnitude of the re-synthesized signal $|\text{STFT}( \text{iSTFT}(\hat{M}e^{j\angle Y}))|$ would not be as good as $\hat{M}$.
This is likely the reason why it is observed in \cite{WPD2019} that extracting ASR features directly from estimated magnitudes, rather than from re-synthesized signals by using the mixture phase, produces better ASR results.

An alternative formulation of MSA uses the RI model of (\ref{ri+istft+magloss}), but uses a weight of zero on the time-domain loss \cite{Kolbak2020loss}:
\begin{align}\label{(ri+istft)x0+magloss}
\mathcal{L}_{\text{(RI-iSTFT)}\times\text{0+Mag}} &= \big\| |\text{STFT}\big(\text{iSTFT}(\hat{S})\big)| - |S| \big\|_1.
\end{align}
We can also do this for the time-domain model in (\ref{wav+magloss}) \cite{Luo2019Fasnet}:
\begin{align}\label{wavx0+magloss}
\mathcal{L}_{\text{Wav}\times 0\text{+Mag}} &= \| |\text{STFT}(\hat{s})| - |S| \|_1.
\end{align}
Essentially, the models are trained to produce a time-domain signal with a good magnitude.
This signal is likely to have a phase not strictly aligned with the clean phase \cite{Kolbak2020loss}, hence would have worse SI-SDR, but would still have reasonable PESQ, STOI, and WER scores, as its magnitude is good.
Such a signal could still be good for human listening \cite{Kolbak2020loss}, as the human auditory system is not sensitive to slight signal shift.

\subsection{Phase Spectrogram Approximation}\label{phase}

The previous sections assume that phase cannot be estimated perfectly.
Magnitude also cannot be estimated perfectly.
In cases where only phase estimates are needed, one could avoid the influence of inaccurate magnitude estimates %
by supplying oracle magnitudes and define a phase loss such as
\begin{align}\label{phaseoraclemagnitude}
\mathcal{L}_{\text{Phase}} &= \| \text{Real}(|S|e^{j\angle (\hat{R}+j\hat{I})}) - \text{Real}(S)\|_1 \nonumber \\
&+ \| \text{Imag}(|S|e^{j\angle (\hat{R}+j\hat{I})}) - \text{Imag}(S)\|_1.
\end{align}

\section{Experimental Setup}\label{setup}

\noindent We validate our ideas on monaural speech enhancement, speaker separation, and ASR tasks.
This section describes the datasets, system configurations, and evaluation metrics.

For speech enhancement, we use the WHAMR!~corpus \cite{Maciejewski2020}, designed for noisy-reverberant 2-speaker separation. %
We tailor the task to noisy-reverberant speech enhancement by removing the second speaker in each mixture.
We use the \textit{min} and 16 kHz version of the corpus, and the first channel for training and evaluation.
We use the target direct sound as the reference for training and metric computation, and perform joint dereverberation and denoising.

For speaker separation, we use the SMS-WSJ dataset \cite{Drude2019}, sampled at 8 kHz.
It contains simulated reverberant 2-speaker mixtures.
The first channel is used for training and evaluation.
We use the direct sound as the training target and perform joint dereverberation, denoising, and separation.
For ASR, we use the default Kaldi-based backend trained on single-speaker noisy-reverberant speech data plus its corresponding direct sound data.
The window length (WL) and hop length (HL) for ASR feature extraction are 25 ms and 10 ms.

For STFT, our separation models use regular 32/8 ms WL/HL for WHAMR!, but 25/10 ms WL/HL for SMS-WSJ to align with the ASR backend. 
For complex spectral mapping and MSA, we use the DenseUNet-TCN architecture
\cite{Wang2020d}. %
For MSA, we use $|Y|$ as features to directly predict $|S|$.
\ZQHL{The same architecture is also used for phase-sensitive spectrogram approximation (PSA) \cite{Erdogan2015}.
The feature is $|Y|$ and the loss is
\begin{align}\label{PSA}
\mathcal{L}_{\text{PSA}} &= \| \hat{M} - |S|\text{\textTau}_0^1 \big(\text{cos}(\angle S-\angle Y)\big)\|_1,
\end{align}
where $\text{\textTau}_0^1(\cdot)$ truncates the values to the range $[0,1]$.}
For time-domain approaches, we employ Conv-TasNet \cite{Luo2019}, where the WL/HL are 5/2.5 ms.
For speaker separation, the loss functions follow (\ref{riloss})-(\ref{PSA}), but we additionally use permutation free (a.k.a.\ invariant) training \cite{Hershey2016,Isik2016, Kolbak2017}.

Our evaluation metrics include SI-SDR \cite{LeRoux2018a}, eSTOI \cite{Taal2011, Manuel2020}, PESQ \cite{Rix2001, Ludlows2020}, and word error rates (WER).
\ZQHL{We point out that SI-SDR is very sensitive to signal shift, while the other measures are not, as magnitude is not sensitive to slight signal shifts.}
Additionally, we use magnitude SNR (mSNR) \cite{Isik2016} to measure the quality of estimated magnitude,
\begin{align}\label{magnitudeSNR}
\small
\text{mSNR} = 10\,\text{log}_{10}\frac{\sum_{t,f} |S(t,f)|^2}{\sum_{t,f} \big| |S(t,f)| - |\hat{S}(t,f)| \big|^2},
\end{align}
and phase SNR (pSNR) to measure that of estimated phase,
\begin{align}\label{phaseSNR}
\small
\text{pSNR} = 10\,\text{log}_{10}\frac{\sum_{t,f} |S(t,f)|^2}{\sum_{t,f} \big| S(t,f) - |S(t,f)|e^{j\angle \hat{S}(t,f)} \big|^2},
\end{align}
where oracle magnitude is supplied for metric computation.

\section{Evaluation Results}\label{results}

\noindent Table~\ref{resultswhamr!} reports the results on WHAMR!'s enhancement task.
The magnitude loss in RI-iSTFT+Mag, \ZQHL{Mag+RI-iSTFT}, and Wav+Mag is computed using 32/8 ms WL/HL.
Comparing RI and RI+Mag, RI-iSTFT and RI-iSTFT+Mag, \ZQHL{RI-iSTFT and Mag+RI-iSTFT}, and Wav and Wav+Mag, we observe clear improvement in PESQ and eSTOI, better mSNR, and slight degradation in SI-SDR, when a magnitude-domain loss is included and signal re-synthesis is performed.
This observation indicates that the estimated magnitude would implicitly compensate for the inaccurate phase estimate when the loss is defined only in the complex- or time-domain, and adding a loss on magnitude would alleviate such compensation.
\ZQHL{With signal re-synthesis, Mag+RI-iSTFT shows worse performance than RI-iSTFT+Mag.}
MSA with re-synthesis shows worse SI-SDR than complex-domain models, likely because MSA uses the mixture phase for re-synthesis.
However, MSA without re-synthesis obtains the best mSNR at 10.9 dB.
This suggests that to obtain a good magnitude, one can consider avoiding modelling magnitude and phase at the same time.
Similarly, the Phase model trained with (\ref{phaseoraclemagnitude}) obtains the best pSNR at 12.4 dB.
Adding a magnitude loss degrades pSNR (for example 10.8 vs.\ 10.4 dB for RI vs.\ RI+Mag).
This indicates the improvement over the mixture phase comes from using a strong DNN model for direct complex- or time-domain prediction, rather than from adding the magnitude loss, a key point that is not discussed in \cite{Pandey2019}.
By only optimizing the magnitude loss, (RI-iSTFT)$\times$0+Mag shows better PESQ, eSTOI and mSNR than RI-iSTFT+Mag, and worse but still good SI-SDR and pSNR scores, even though no time-domain loss is included.
This indicates the complex-domain model implicitly figures out a reasonably-good phase that can produce a good magnitude.
(RI-iSTFT)$\times$0+Mag produces worse mSNR than MSA (12.9 vs.\ 13.1 dB), possibly because the magnitude has to be extracted from a time-domain signal, which may limit the model's capability at magnitude estimation.
Wav$\times$0+Mag obtains good mSNR, PESQ, and eSTOI, but much worse SI-SDR and pSNR scores.
This degradation might be due to the small WL in Conv-TasNet.
The Wav$\times$0+Mag results provide a strong experimental evidence showing that PESQ and eSTOI scores largely depend on the magnitude of the estimated signal.

\begin{table}[t]
\centering
 \sisetup{table-format=2.2,round-mode=places,round-precision=2,table-number-alignment = center,detect-weight=true,detect-inline-weight=math}
\caption{\scriptsize SI-SDR (dB), PESQ, eSTOI (\%), mSNR (dB), and pSNR (dB) on WHAMR! (Enh.)} %
\vspace{-0.15cm}
\label{resultswhamr!}
\setlength{\tabcolsep}{2.5pt}
\resizebox{.98\linewidth}{!}
{
\begin{tabular}{l|c|c|c|S[table-format=2.1,round-precision=1]S[table-format=1.2]S[table-format=2.1,round-precision=1]S[table-format=2.1,round-precision=1]S[table-format=2.1,round-precision=1]}
\toprule
& & WL/HL & Signal & \multicolumn{1}{c}{SI-} & & & & \\
Approaches & Eq. & (ms) & re-syn? & \multicolumn{1}{c}{SDR} & \multicolumn{1}{c}{PESQ} & \multicolumn{1}{c}{eSTOI} & \multicolumn{1}{c}{mSNR} & \multicolumn{1}{c}{pSNR} \\ \midrule
Unprocessed & {-} & 32/8 & - & -2.7 & 1.53 & 45.1 & -1.63 & -2.8 \\

\midrule

MSA & (\ref{MSA}) & 32/8 & yes & 4.4 & 2.72 & 78.5 & 11.34 & 6.29 \\
MSA & (\ref{MSA}) & 32/8 & no & \multicolumn{1}{c}{-} & \multicolumn{1}{c}{-} & \multicolumn{1}{c}{-} & 13.05 & \multicolumn{1}{c}{-}\\
\midrule

RI & (\ref{riloss}) & 32/8 & yes & 9.1 & 2.49 & 80.3 & 12.66 & 10.8 \\
RI & (\ref{riloss}) & 32/8 & no & \multicolumn{1}{c}{-} & \multicolumn{1}{c}{-} & \multicolumn{1}{c}{-} & 12.58 & 10.751 \\
RI+Mag & (\ref{ri+magloss}) & 32/8 & yes & 8.6 & 2.92 & 81.9 & 12.84 & 10.35 \\
RI+Mag & (\ref{ri+magloss}) & 32/8 & no & \multicolumn{1}{c}{-} & \multicolumn{1}{c}{-} & \multicolumn{1}{c}{-} & 12.82 & 10.24 \\
RI-iSTFT & (\ref{ri-istftloss}) & 32/8 & yes & 8.80 & 2.46 & 79.0 & 12.08 & 10.79 \\
Mag+RI-iSTFT & (\ref{magloss+ri+istft}) & 32/8 & yes & 8.4 & 2.55 & 80.9 & 12.5 & 10.1 \\
Mag+RI-iSTFT & (\ref{magloss+ri+istft}) & 32/8 & no & {-} & {-} & {-} & 12.9 & 9.4 \\
RI-iSTFT+Mag & (\ref{ri+istft+magloss}) & 32/8 & yes & 8.56 & 2.86 & 81.7 & 12.64 & 10.46 \\
(RI-iSTFT)$\times$0+Mag & (\ref{(ri+istft)x0+magloss}) & 32/8 & yes & 7.3 & 2.91 & 82.7 & 12.9 & 9.1 \\
Phase & (\ref{phaseoraclemagnitude}) & 32/8 & no & \multicolumn{1}{c}{-} & \multicolumn{1}{c}{-} & \multicolumn{1}{c}{-} & \multicolumn{1}{c}{-} & 12.4 \\
\midrule

Wav & (\ref{wavloss}) & 5/2.5 & yes & 7.7 & 2.20 & 78.0 & 11.22 & 10.14 \\
Wav+Mag & (\ref{wav+magloss}) & 5/2.5 & yes & 7.5 & 2.58 & 80.1 & 11.43 & 9.90 \\
Wav$\times$0+Mag & (\ref{wavx0+magloss}) & 5/2.5 & yes & -9.09 & 2.67 & 80.6 & 11.36 & -3.64 \\

\midrule

PSA & (\ref{PSA}) & 32/8 & yes & 5.6 & 2.36 & 76.3 & 10.2 & 7.2 \\
PSA & (\ref{PSA}) & 32/8 & no & \multicolumn{1}{c}{-} & \multicolumn{1}{c}{-} & \multicolumn{1}{c}{-} & 9.9 & \multicolumn{1}{c}{-} \\

\midrule

PSM & {-} & 32/8 & yes & 8.43 & 3.82 & 91.1 & 13.40 & 9.63 \\ 
IAM & {-} & 32/8 & yes & 5.37 & 3.47 & 89.7 & 14.85 & 6.52 \\

\bottomrule
\end{tabular}
}
\vspace{-0.4cm}
\end{table}

\begin{figure}
  \centering

  \includegraphics[width=9cm]{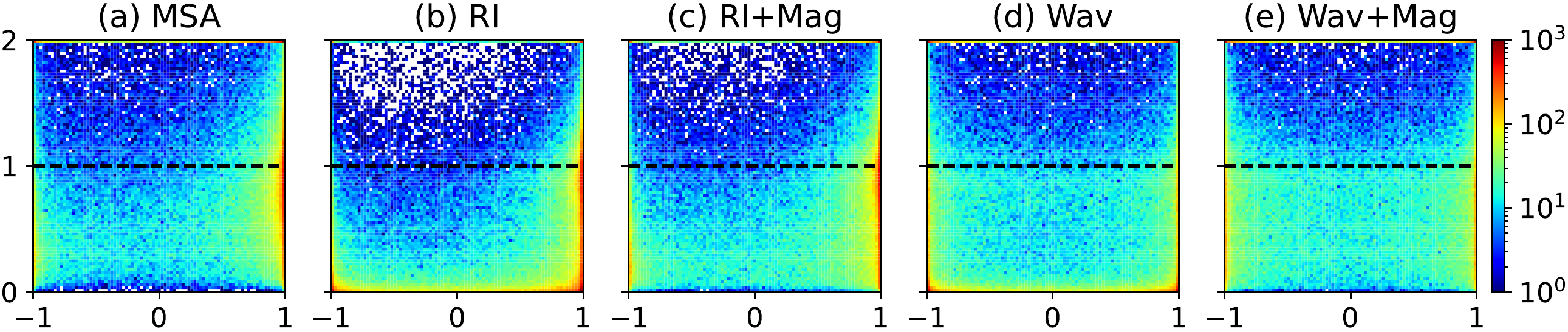}

\vspace{-0.1cm}
  \caption{
  \scriptsize
  2D histograms for various models.
  The x-axis is the phase difference $\text{cos}(\angle S(t,f)-\angle Y(t,f))$,
  which gets smaller as $\angle Y(t,f)$ gets away from $\angle S(t,f)$,
  and the y-axis is the magnitude ratio $\hat{M}(t,f)/|S(t,f)|$ (truncated to the range $[0,2]$). %
  For RI and RI+Mag, $\hat{M}=|\hat{R}+j\hat{I}|$ (i.e., no re-synthesis).
  For Wav and Wav+Mag, $\hat{M}$ is extracted from the re-synthesized signal.
  In this example, the mixture mSNR is -2.3 dB, and the output mSNRs are 12.4, 11.0, 12.1, 10.7 and 11.3 dB for the MSA, RI, RI+Mag, WA and WA+Mag models, respectively.
  Before plotting, we throw away T-F units whose energy in $|S|$ is more than 60 dB weaker than the highest-energy T-F unit.
  }
  \vspace{-0.5cm}
  \label{scatterplot}
\end{figure}

To illustrate the compensation of estimated magnitudes when mixture phase $\angle Y(t,f)$ is different from clean phase $\angle S(t,f)$, we provide two-dimensional (2D) histograms of phase difference vs.\ magnitude ratios in Fig.~\ref{scatterplot}, based on a test mixture of WHAMR!.
See the caption for the definitions of the axes and other details.
Comparing MSA and RI, we observe that the
magnitude ratios by MSA are overall closer to the dashed line, which denotes the case of perfect magnitude estimation.
This indicates that MSA produces better magnitude estimation.
In addition, for RI, many estimated magnitudes are much smaller than the clean magnitudes (i.e., near the bottom of the plot), while the magnitude ratios in MSA are much less dense near the bottom.
This suggests that when $\angle Y(t,f)$ is different from $\angle S(t,f)$ and hence $\angle \hat{S}(t,f)$ likely differs from $\angle S(t,f)$, RI compresses (or sacrifices) the estimated magnitude to better approximate $S(t,f)$, while MSA does not because MSA assumes that $\angle \hat{S}(t,f)=\angle S(t,f)$.
Adding a magnitude loss to RI, RI+Mag improves the magnitude estimation, but still produces less accurate magnitude than MSA, as the magnitude ratios are more spread out than MSA when $\angle Y(t,f)$ is different from $\angle S(t,f)$.
Similar trends are observed in the WA and WA+Mag plots.
These histograms indicate the validity of the magnitude-phase compensation phenomenon, and that direct MSA produces better magnitudes.

Table~\ref{resultssmswsj} reports the results on SMS-WSJ.
In this setup, the magnitude loss in RI-iSTFT+Mag and Wav+Mag is computed based on 25/10 ms WL/HL.
Adding a magnitude loss leads to clearly better WER for RI, RI-iSTFT, and Wav.
MSA exhibits strong WER with no re-synthesis at 33.87\%, and competitive WER with re-synthesis at 32.84\%, possibly because $\angle Y$ used for re-synthesis does not dramatically degrade the magnitude. 
The trend on the other metrics is similar to that in Table~\ref{resultswhamr!}.

In both tables, we report oracle real-valued T-F masking scores of IAM ($|S|/|Y|$) \cite{WYXtrainingtargets} and PSM \cite{Erdogan2015}, both using $\angle Y$ for re-synthesis.
IAM makes an aggressive step $|S(t,f)|$ along $\angle Y(t,f)$, while PSM makes a less aggressive step $|S(t,f)|\text{cos}(\angle S(t,f)-\angle Y(t,f))$, considering that $\angle Y(t,f)$ is different from $\angle S(t,f)$.
PSM shows better SI-SDR, as the masked mixture spectrum is closer to the clean one.
IAM with re-synthesis shows better mSNR and WER than PSM, likely because the magnitude used for re-synthesis is oracle and the re-synthesized signal still has a reasonable magnitude, even though $\angle Y$ is used for re-synthesis.
\ZQHL{Comparing MSA and PSA, we observe that PSA obtains better SI-SDR while worse scores on the other metrics that favor a good magnitude. The PSA model shows worse performance than end-to-end models.}

\begin{table}[t]
\centering
 \sisetup{table-format=2.2,round-mode=places,round-precision=2,table-number-alignment = center,detect-weight=true,detect-inline-weight=math}
\caption{\scriptsize SI-SDR (dB), PESQ, eSTOI (\%), mSNR (dB), and WER (\%) %
on SMS-WSJ.}
\vspace{-0.15cm}
\label{resultssmswsj}
\setlength{\tabcolsep}{2.5pt}
\resizebox{.98\linewidth}{!}
{
\begin{tabular}{l|
c|
c|
c|
S[table-format=2.1,round-precision=1]
S[table-format=1.2]
S[table-format=2.1,round-precision=1]
S[table-format=2.1,round-precision=1]
S}
\toprule
 & & WL/HL & Signal & \multicolumn{1}{c}{SI-} & & & & \\ 
Approaches & Eq. & (ms) & re-syn? & \multicolumn{1}{c}{SDR} & \multicolumn{1}{c}{PESQ} & \multicolumn{1}{c}{eSTOI} & \multicolumn{1}{c}{mSNR} & \multicolumn{1}{c}{WER} \\ 
\midrule
Unprocessed & {-} & 25/10 & - & -5.5 & 1.50 & 44.1 & -4.44 & 79.43 \\
\midrule

MSA & (\ref{MSA}) & 25/10 &yes & 0.25 & 2.20 & 69.8 & 8.42 & 32.84 \\ 
MSA & (\ref{MSA}) & 25/10 & no & \multicolumn{1}{c}{-} & \multicolumn{1}{c}{-} & \multicolumn{1}{c}{-} & 9.57 & 33.87 \\
\midrule

RI & (\ref{riloss}) & 25/10 & yes & 4.64499 & 1.9746 & 70.089 & 8.7227 & 42.26 \\ 
RI & (\ref{riloss}) & 25/10 & no & \multicolumn{1}{c}{-} & \multicolumn{1}{c}{-} & \multicolumn{1}{c}{-} & 8.6307 & 40.98 \\ 
RI+Mag & (\ref{ri+magloss}) & 25/10 & yes & 3.23 & 2.21 & 70.1 & 8.59 & 35.97 \\ 
RI+Mag & (\ref{ri+magloss}) & 25/10 & no & \multicolumn{1}{c}{-} & \multicolumn{1}{c}{-} & \multicolumn{1}{c}{-} & 8.61 & 36.05 \\
RI-iSTFT & (\ref{ri-istftloss}) & 25/10 & yes & 4.51 & 1.87 & 67.8 & 8.00 & 44.58 \\ 
Mag+RI-iSTFT & (\ref{magloss+ri+istft}) & 25/10 & yes & 3.2 & 1.99 & 69.4 & 8.3 & 39.09 \\
Mag+RI-iSTFT & (\ref{magloss+ri+istft}) & 25/10 & no & {-} & {-} & {-} & 9.5 & 33.98 \\
RI-iSTFT+Mag & (\ref{ri+istft+magloss}) & 25/10 & yes & 3.457 & 2.24 & 71.3 & 9.008 & 33.66 \\ 
(RI-iSTFT)$\times$0+Mag & (\ref{(ri+istft)x0+magloss}) & 25/10 & yes & 1.96 & 2.2895 & 72.076 & 9.04065 & 32.38 \\ 

\midrule

Wav & (\ref{wavloss}) & 5/2.5 & yes & 4.22 & 1.79 & 66.3 & 8.33 & 47.50 \\ 
Wav+Mag & (\ref{wav+magloss}) & 5/2.5 & yes & 3.42 & 2.07 & 68.9 & 8.33 & 39.22 \\
Wav$\times$0+Mag & (\ref{wavx0+magloss}) & 5/2.5 & yes & -4.1528 & 2.11 & 69.58 & 8.24 & 37.91 \\
\midrule

PSA & (\ref{PSA}) & 25/10 & yes & 1.4 & 1.82 & 65.6 & 6.7 & 44.75 \\ 
PSA & (\ref{PSA}) & 25/10 & no & \multicolumn{1}{c}{-} & \multicolumn{1}{c}{-} & \multicolumn{1}{c}{-} & 6.5 & 48.15 \\
\midrule

PSM & {-} & 25/10 & yes & 5.79 & 3.64 & 89.7 & 10.46 & 5.84 \\
IAM & {-} & 25/10 & yes & 1.53 & 3.37 & 91.1 & 13.11 & 5.71 \\
IAM & {-} & 25/10 & no & \multicolumn{1}{c}{-} & \multicolumn{1}{c}{-} & \multicolumn{1}{c}{-} & $\infty$ & 5.46 \\
\bottomrule
\end{tabular}
}
\vspace{-0.6cm}
\end{table}

\section{Conclusion}\label{conclusion}

\noindent We have provided a novel view on the implicit compensation between estimated magnitude and phase in DNN based speech separation.
This view provides a fundamental understanding of the performance differences between including and not including a magnitude-domain loss for training.
This understanding can benefit the design of many source separation algorithms and has broad implications.
For example, when using time-domain models as benchmarks and PESQ, STOI, or WER as the evaluation metrics, a study should consider training its time-domain models with a magnitude-domain loss in combination with a time-domain loss.
Another such example is monaural sound event detection that performs separation before detection, where many studies train the separator to optimize SI-SDR \cite{Turpault2020, Wisdom2021}.
One could train the separator by including a magnitude loss, as the detector is usually trained on energy features, \ZQHL{or train the separator jointly with the detector to only optimize the detection loss, similarly to the joint frontend and backend training \cite{Wang2016,Chang2019} in robust ASR.}

\bibliographystyle{IEEEtran}
\bibliography{references.bib}

\begin{thebibliography}{10}
\providecommand{\url}[1]{#1}
\csname url@samestyle\endcsname
\providecommand{\newblock}{\relax}
\providecommand{\bibinfo}[2]{#2}
\providecommand{\BIBentrySTDinterwordspacing}{\spaceskip=0pt\relax}
\providecommand{\BIBentryALTinterwordstretchfactor}{4}
\providecommand{\BIBentryALTinterwordspacing}{\spaceskip=\fontdimen2\font plus
\BIBentryALTinterwordstretchfactor\fontdimen3\font minus
  \fontdimen4\font\relax}
\providecommand{\BIBforeignlanguage}[2]{{%
\expandafter\ifx\csname l@#1\endcsname\relax
\typeout{** WARNING: IEEEtran.bst: No hyphenation pattern has been}%
\typeout{** loaded for the language `#1'. Using the pattern for}%
\typeout{** the default language instead.}%
\else
\language=\csname l@#1\endcsname
\fi
#2}}
\providecommand{\BIBdecl}{\relax}
\BIBdecl

\bibitem{WDLreview}
D.~Wang and J.~Chen, ``{Supervised Speech Separation Based on Deep Learning: An
  Overview},'' \emph{IEEE/ACM Trans. Audio, Speech, Lang. Process.}, vol.~26,
  pp. 1702--1726, 2018.

\bibitem{WYXtrainingtargets}
Y.~Wang, A.~Narayanan, and D.~Wang, ``{On Training Targets for Supervised
  Speech Separation},'' \emph{IEEE/ACM Trans. Audio, Speech, Lang. Process.},
  vol.~22, pp. 1849--1858, 2014.

\bibitem{Erdogan2015}
H.~Erdogan, J.~R. Hershey, S.~Watanabe, and J.~{Le Roux}, ``{Phase-Sensitive
  and Recognition-Boosted Speech Separation using Deep Recurrent Neural
  Networks},'' in \emph{Proc. ICASSP}, 2015, pp. 708--712.

\bibitem{Han2015}
K.~Han, Y.~Wang, D.~Wang, W.~{S. Woods}, I.~Merks, and T.~Zhang, ``{Learning
  Spectral Mapping for Speech Dereverberation and Denoising},'' \emph{IEEE
  Trans. Audio, Speech, Lang. Process.}, vol.~23, no.~6, pp. 982--992, 2015.

\bibitem{WZQe2eMISI2018}
Z.-Q. Wang, J.~{Le Roux}, D.~Wang, and J.~R. Hershey, ``{End-to-End Speech
  Separation with Unfolded Iterative Phase Reconstruction},'' in \emph{Proc.
  Interspeech}, 2018, pp. 2708--2712.

\bibitem{WZQtrigonometric2019}
Z.-Q. Wang, K.~Tan, and D.~Wang, ``{Deep Learning Based Phase Reconstruction
  for Speaker Separation: A Trigonometric Perspective},'' in \emph{Proc.
  ICASSP}, 2019, pp. 71--75.

\bibitem{Zhao2019}
Y.~Zhao, Z.-Q. Wang, and D.~Wang, ``{Two-Stage Deep Learning for
  Noisy-Reverberant Speech Enhancement},'' \emph{IEEE/ACM Trans. Audio, Speech,
  Lang. Process.}, vol.~27, no.~1, pp. 53--62, 2019.

\bibitem{Williamson2016}
D.~S. Williamson, Y.~Wang, and D.~Wang, ``{Complex Ratio Masking for Monaural
  Speech Separation},'' \emph{IEEE/ACM Trans. Audio, Speech, Lang. Process.},
  pp. 483--492, 2016.

\bibitem{Fu2017}
S.-W. Fu, T.-Y. Hu, Y.~Tsao, and X.~Lu, ``{Complex Spectrogram Enhancement By
  Convolutional Neural Network with Multi-Metrics Learning},'' in \emph{Proc.
  MLSP}, 2017, pp. 1--6.

\bibitem{tan2020}
K.~Tan and D.~Wang, ``{Learning Complex Spectral Mapping With Gated
  Convolutional Recurrent Networks for Monaural Speech Enhancement},''
  \emph{IEEE/ACM Trans. Audio, Speech, Lang. Process.}, vol.~28, pp. 380--390,
  2020.

\bibitem{Wang2020a}
Z.-Q. Wang, P.~Wang, and D.~Wang, ``{Complex Spectral Mapping for Single-and
  Multi-Channel Speech Enhancement and Robust ASR},'' \emph{IEEE/ACM Trans.
  Audio, Speech, Lang. Process.}, vol.~28, pp. 1778--1787, 2020.

\bibitem{Wang2020b}
Z.-Q. Wang and D.~Wang, ``{Deep Learning Based Target Cancellation for Speech
  Dereverberation},'' \emph{IEEE/ACM Trans. Audio, Speech, Lang. Process.},
  vol.~28, pp. 941--950, 2020.

\bibitem{Wang2020d}
------, ``{Multi-Microphone Complex Spectral Mapping for Speech
  Dereverberation},'' in \emph{Proc. ICASSP}, 2020, pp. 486--490.

\bibitem{Wang2020c}
Z.-Q. Wang, P.~Wang, and D.~Wang, ``{Multi-Microphone Complex Spectral Mapping
  for Utterance-Wise and Continuous Speaker Separation},'' \emph{IEEE/ACM
  Trans. Audio, Speech, Lang. Process.}, 2021.

\bibitem{Pascual2017}
S.~Pascual, A.~Bonafonte, and J.~Serr, ``{SEGAN : Speech Enhancement Generative
  Adversarial Network},'' in \emph{Proc. Interspeech}, 2017.

\bibitem{Rethage2018}
D.~Rethage, J.~Pons, and X.~Serra, ``{A WaveNet for Speech Denoising},'' in
  \emph{Proc. ICASSP}, 2018, pp. 5069--5073.

\bibitem{Luo2019}
Y.~Luo and N.~Mesgarani, ``{Conv-TasNet: Surpassing Ideal Time-Frequency
  Magnitude Masking for Speech Separation},'' \emph{IEEE/ACM Trans. Audio,
  Speech, Lang. Process.}, vol.~27, no.~8, pp. 1256--1266, 2019.

\bibitem{Pandey2019}
A.~Pandey and D.~Wang, ``{A New Framework for CNN-Based Speech Enhancement in
  the Time Domain},'' \emph{IEEE/ACM Trans. Audio, Speech, Lang. Process.},
  vol.~27, pp. 1179--1188, 2019.

\bibitem{Defossez2020}
A.~D{\'{e}}fossez, G.~Synnaeve, and Y.~Adi, ``{Real Time Speech Enhancement in
  the Waveform Domain},'' in \emph{Proc. Interspeech}, 2020.

\bibitem{Liu2019}
Y.~Liu and D.~Wang, ``{Divide and Conquer: A Deep CASA Approach to
  Talker-Independent Monaural Speaker Separation},'' \emph{IEEE/ACM Trans.
  Audio, Speech, Lang. Process.}, vol.~27, no.~12, pp. 2092--2102, 2019.

\bibitem{Isik2020}
U.~Isik, R.~Giri, N.~Phansalkar, J.~M. Valin, K.~Helwani, and A.~Krishnaswamy,
  ``{PoCoNet: Better Speech Enhancement with Frequency-Positional Embeddings,
  Semi-Supervised Conversational Data, and Biased Loss},'' in \emph{Proc.
  Interspeech}, 2020, pp. 2487--2491.

\bibitem{Wisdom2018}
S.~Wisdom, J.~R. Hershey, K.~Wilson, J.~Thorpe, M.~Chinen, B.~Patton, and R.~A.
  Saurous, ``{Differentiable Consistency Constraints for Improved Deep Speech
  Enhancement},'' in \emph{Proc. ICASSP}, 2019, pp. 900--904.

\bibitem{Li2021}
A.~Li, W.~Liu, C.~Zheng, C.~Fan, and X.~Li, ``{Two Heads Are Better Than One: A
  Two-Stage Complex Spectral Mapping Approach for Monaural Speech
  Enhancement},'' \emph{IEEE/ACM Trans. Audio, Speech, Lang. Process.}, 2021.

\bibitem{Braun2021}
S.~Braun, H.~Gamper, C.~{K. A. Reddy}, and I.~Tashev, ``{Towards Efficient
  Models for Real-Time Deep Noise Suppression},'' in \emph{Proc. ICASSP}, 2021,
  pp. 656--660.

\bibitem{Aroudi2021}
A.~Aroudi and S.~Braun, ``{DBnet: Doa-Driven Beamforming Network for End-to-End
  Reverberant Sound Source Separation},'' in \emph{Proc. ICASSP}, 2021, pp.
  211--215.

\bibitem{Sawata2021}
R.~Sawata, S.~Uhlich, S.~Takahashi, and Y.~Mitsufuji, ``{All For One And One
  For All: Improving Music Separation By Bridging Networks},'' in \emph{Proc.
  ICASSP}, 2021, pp. 51--55.

\bibitem{Li2021power}
A.~Li, C.~Zheng, R.~Peng, and X.~Li, ``{On The Importance of Power Compression
  and Phase Estimation in Monaural Speech Dereverberation},'' \emph{JASA
  Express Letters}, vol.~1, no.~1, p. 014802, 2021.

\bibitem{G.Germain2019}
F.~{G. Germain}, Q.~Chen, and V.~Koltun, ``{Speech Denoising with Deep Feature
  Losses},'' in \emph{Proc. Interspeech}, 2019, pp. 2723--2727.

\bibitem{Manocha2021}
P.~Manocha, Z.~Jin, R.~Zhang, and A.~Finkelstein, ``{CDPAM: Contrastive
  Learning for Perceptual Audio Similarity},'' in \emph{Proc. ICASSP}, 2021,
  pp. 196--200.

\bibitem{Stoller2018}
D.~Stoller, S.~Ewert, and S.~Dixon, ``{Wave-U-Net: A Multi-Scale Neural Network
  for End-to-End Audio Source Separation},'' in \emph{Proc. ISMIR}, 2018, pp.
  334--340.

\bibitem{LeRoux2018a}
J.~Le~Roux, S.~Wisdom, H.~Erdogan, and J.~R. Hershey, ``{SDR – Half-Baked or
  Well Done?}'' in \emph{Proc. ICASSP}, 2019, pp. 626--630.

\bibitem{Rix2001}
A.~Rix, J.~Beerends, M.~Hollier, and A.~Hekstra, ``{Perceptual Evaluation of
  Speech Quality (PESQ)-A New Method for Speech Quality Assessment of Telephone
  Networks and Codecs},'' in \emph{Proc. ICASSP}, vol.~2, 2001, pp. 749--752.

\bibitem{Taal2011}
C.~H. Taal, R.~C. Hendriks, R.~Heusdens, and J.~Jensen, ``{An Algorithm for
  Intelligibility Prediction of Time–Frequency Weighted Noisy Speech},''
  \emph{IEEE Trans. Audio, Speech, Lang. Process.}, vol.~19, no.~7, pp.
  2125--2136, sep 2011.

\bibitem{WPD2019}
P.~Wang and D.~Wang, ``{Enhanced Spectral Features for Distortion-Independent
  Acoustic Modeling},'' in \emph{Proc. Interspeech}, 2019.

\bibitem{Courville2016}
\BIBentryALTinterwordspacing
A.~Courville, I.~Goodfellow, and Y.~Bengio, \emph{{Deep Learning}}.\hskip 1em
  plus 0.5em minus 0.4em\relax MIT Press, 2016. [Online]. Available:
  \url{http://www.deeplearningbook.org}
\BIBentrySTDinterwordspacing

\bibitem{Griffin1984a}
D.~W. Griffin and J.~S. Lim, ``{Signal Estimation from Modified Short-Time
  Fourier Transform},'' \emph{IEEE Trans. Audio, Speech, Signal Process.},
  vol.~32, no.~2, pp. 236--243, 1984.

\bibitem{LeRoux2008}
J.~{Le Roux}, N.~Ono, and S.~Sagayama, ``{Explicit Consistency Constraints for
  STFT Spectrograms and Their Application to Phase Reconstruction},''
  \emph{Proceedings of SAPA}, 2008.

\bibitem{Kolbak2020loss}
M.~Kolb{\ae}k, Z.-H. Tan, S.~H. Jensen, and J.~Jensen, ``{On Loss Functions for
  Supervised Monaural Time-Domain Speech Enhancement},'' \emph{IEEE/ACM Trans.
  Audio, Speech, Lang. Process.}, vol.~28, pp. 825--838, 2020.

\bibitem{Luo2019Fasnet}
Y.~Luo, E.~Ceolini, C.~Han, S.-C. Liu, and N.~Mesgarani, ``{FaSNet: Low-latency
  Adaptive Beamforming for Multi-Microphone Audio Processing},'' in \emph{Proc.
  WASPAA}, 2019.

\bibitem{Maciejewski2020}
M.~Maciejewski, G.~Wichern, E.~McQuinn, and J.~{Le Roux}, ``{WHAMR!: Noisy and
  Reverberant Single-Channel Speech Separation},'' in \emph{Proc. ICASSP},
  2020.

\bibitem{Drude2019}
L.~Drude, J.~Heitkaemper, C.~Boeddeker, and R.~Haeb-Umbach, ``{SMS-WSJ:
  Database, Performance Measures, and Baseline Recipe for Multi-Channel Source
  Separation and Recognition},'' in \emph{arXiv preprint arXiv:1910.13934},
  2019.

\bibitem{Hershey2016}
J.~R. Hershey, Z.~Chen, J.~{Le Roux}, and S.~Watanabe, ``{Deep Clustering:
  Discriminative Embeddings for Segmentation and Separation},'' in \emph{Proc.
  ICASSP}, 2016, pp. 31--35.

\bibitem{Isik2016}
Y.~Isik, J.~{Le Roux}, Z.~Chen, S.~Watanabe, and J.~R. Hershey,
  ``Single-channel multi-speaker separation using deep clustering,'' in
  \emph{Proc. Interspeech}, Sep. 2016, pp. 545--549.

\bibitem{Kolbak2017}
M.~Kolb{\ae}k, D.~Yu, Z.-H. Tan, and J.~Jensen, ``{Multi-Talker Speech
  Separation with Utterance-Level Permutation Invariant Training of Deep
  Recurrent Neural Networks},'' \emph{IEEE/ACM Trans. Audio, Speech, Lang.
  Process.}, vol.~25, no.~10, pp. 1901--1913, 2017.

\bibitem{Manuel2020}
P.~Manuel, ``https://github.com/mpariente/pystoi,'' 2020.

\bibitem{Ludlows2020}
Ludlows, ``https://github.com/ludlows/python-pesq,'' 2020.

\bibitem{Turpault2020}
N.~Turpault, S.~Wisdom, H.~Erdogan, J.~Hershey, R.~Serizel, E.~Fonseca,
  P.~Seetharaman, and J.~Salamon, ``{Improving Sound Event Detection in
  Domestic Environments using Sound Separation},'' in \emph{Proc. DCASE}, 2020.

\bibitem{Wisdom2021}
S.~Wisdom, H.~Erdogan, D.~{P. W. Ellis}, R.~Serizel, N.~Turpault, E.~Fonseca,
  J.~Salamon, P.~Seetharaman, and J.~{R. Hershey}, ``{What's All The Fuss About
  Free Universal Sound Separation Data?}'' in \emph{Proc. ICASSP}, 2021, pp.
  186--190.

\bibitem{Wang2016}
Z.-Q. Wang and D.~Wang, ``{A Joint Training Framework for Robust Automatic
  Speech Recognition},'' \emph{IEEE/ACM Trans. Audio, Speech, Lang. Process.},
  vol.~24, no.~4, pp. 796--806, 2016.

\bibitem{Chang2019}
X.~Chang, W.~Zhang, Y.~Qian, J.~{Le Roux}, and S.~Watanabe, ``{MIMO-SPEECH:
  End-to-End Multi-Channel Multi-Speaker Speech Recognition},'' in \emph{Proc.
  ASRU}, 2019.

\end{thebibliography}

\end{document}